\documentclass[arxiv,usenatbib]{iupartex}
\usepackage{newtxtext,newtxmath}
\usepackage[T1]{fontenc}

\title[UBV Transformation for TÜBİTAK T100 Telescope]{Transformation relations for UBV photometric system of 1m telescope at the T\"{U}B\.{I}TAK National Observatory}

\author[Ak et. al]{%
T. Ak$^{1\cc}$,\orcid{0000-0002-0688-1983}
R. Canbay$^{2}$,\orcid{00000-0003-2575-9892}
and
T. Yontan $^{1}$\orcid{0000-0002-5657-6194}
\affsep \\
$^1$Istanbul University, Faculty of Science, Department of Astronomy and Space Sciences, 34119, Istanbul, T\"{u}rkiye\\
$^2$Istanbul University, Institute of Graduate Studies in Science, Programme of Astronomy and Space Sciences, 34116, Istanbul, T\"{u}rkiye\\
}

\corres{T. Ak}{tanselak@istanbul.edu.tr}

\pubyear{20XX}

\doiheader{XXXXXXX/PAR.20XX.00000}
\date{
	\pSubmit{01.11.2024} 
	\pAccept{30.11.2024}
	\pPubOnl{00.00.0000}
}
\volume{0}
\volnumber{1}

\begin{document}
\label{firstpage}
\pagerange{\pageref*{firstpage}--\pageref*{lastpage}}
\maketitle

\begin{abstract}
$UBV$ CCD observations of standard stars selected from \citet{Land09, Land13} were performed using the 1-meter telescope (T100) of the T\"{U}B\.{I}TAK National Observatory equipped with a back-illuminated and UV enhanced CCD camera and Bessell $UBV$ filters. Observations were conducted over an extended period, spanning from 2012 to 2024, covering a total of 50 photometric nights. Photometric measurements were used to find the standard transformation relations of the T100 photometric system. The atmospheric extinction coefficients, zero points and transformation coefficients of each night were determined. No time dependence was found for the secondary extinction coefficients. However, it was observed that the primary extinction coefficients decreased until the year 2019 and increased after that year. Strong seasonal variations in the extinction coefficients were not evident. Small differences in seasonal median values of them were used to attempt to find the atmospheric extinction sources. We found calculated minus catalogue values for each standard star, $\Delta(U-B)$, $\Delta(B-V)$ and $\Delta V$. The means and standard deviations of these differences were estimated to be 1.4$\pm$76, 1.9$\pm$18 and 0.0$\pm$36 mmag, respectively. We found that our data well matched Landolt's standards for $V$ and $B-V$, i.e. there are no systematic differences. However, there are systematic differences for $U-B$ between the two photometric systems, which is probably originated from the quantum efficiency differences of the detectors used in the photometric systems, although the median differences are relatively small ($|\Delta(U-B)|$< 50 mmag) for stars with $-0.5<U-B~{(\rm mag)} <1.6$ and $0.2<B-V~{(\rm mag)} <1.8$. As an overall result, we conclude that the transformation relations found in this study can be used for standardized photometry with the T100 photometric system. 
\end{abstract}


%%%%%%%%%%%%%%%%%%%%%%%%%%%%%%%%%%%%%%%%%%%%%%%%%%

%%%%%%%%%%%%%%%%% BODY OF PAPER %%%%%%%%%%%%%%%%%%

\section{Introduction}
The instrumental magnitude of a celestial object measured during an astronomical observation depends not only on the object’s flux and atmospheric extinction but also on the spectral response and transmission properties of the telescope-filter-detector combination. In some cases, instrumental magnitudes must be transformed into a standard photometric system by observing standard stars. The ‘Johnson-Kron-Cousins’ $UBVR_{\rm C}I_{\rm C}$ system is the most widely used broad-band photometric system. The $UBV$($RI$) photometric system was designed by \citet{Johnson1953} taking Yerkes Atlas system (MK) of spectral classification as standard. In this photometric system, the colour indices of the bright star Vega with spectral type A0 was defined as the zero point of all colour indices. Due to advancements in detector technologies, accurate photometry of faint stars became possible in the 1970s and 1990s, and the Kron-Cousins $R_{\rm C}I_{\rm C}$ filters were replaced with $RI$ filters of Johnson and Morgan.
 
There are two main sets of standard stars used for broadband $UBVR_{\rm C}I_{\rm C}$ photometry. The fundamental standard stars for $UBVR_{\rm C}I_{\rm C}$ photometry in the southern hemisphere are provided by E-region standards which are centered at declination –45$^{\circ}$. The E-region standard star photometry was established by Dr. A. W. J. Cousins. Its accuracy was refined and extended in colour range by SAAO astronomers \citep{Kilkenny1998, Menzies1989}. For the northern hemisphere, standard stars along the celestial equator \citep[][and references therein]{Land09} and ones centered approximately +50$^{\circ}$ declination \citep{Land13} are used \citep[see also,][]{Menzies1991}. Landolt's standards along the celestial equator can be observed from both hemispheres. Since Landolt extended his standards to faint stars, they can be observed with CCD detectors attached to large telescopes. It is easier to calculate the transformation coefficients using these stars because they include blue and red stars in relatively small areas. Since both systems were established using photoelectric photometers and astronomers use CCD cameras for current photometric observations, transformations to the standard system are necessary. These transformations are generally linear as combined spectral responses of filter and detector are very similar \citep[see,][]{Sung2000}. 

Detailed information of photometric observing systems, including atmospheric extinction coefficients and transformation relations, is crucial for standardized photometry.  In a series of {\it UBV} photometric observations of open stellar clusters between the years 2012 and 2024, we have also observed Landolt’s selected standard star fields for each observing night with Bessell $UBV$ filters attached to an SI 1100S CCD camera and 1-meter telescope of the T\"{U}B\.{I}TAK National Observatory. Atmospheric extinction coefficients and transformation equations to standard photometric systems were calculated for each photometric night. Although the observations have not been done specifically for monitoring the extinction and transformation coefficients of this photometric system, we could obtain them as a side-product of our observations. In this study, we investigate the variation of atmospheric extinction coefficients for the last 12 years and introduce a reliable set of $UBV$ transformation relations for the photometric observing system of the 1-meter telescope (T100) at the T\"{U}B\.{I}TAK National Observatory.  

\section{Observations and data reduction}

\begin{figure}
	\begin{center}
		\includegraphics[width=1.0\columnwidth]{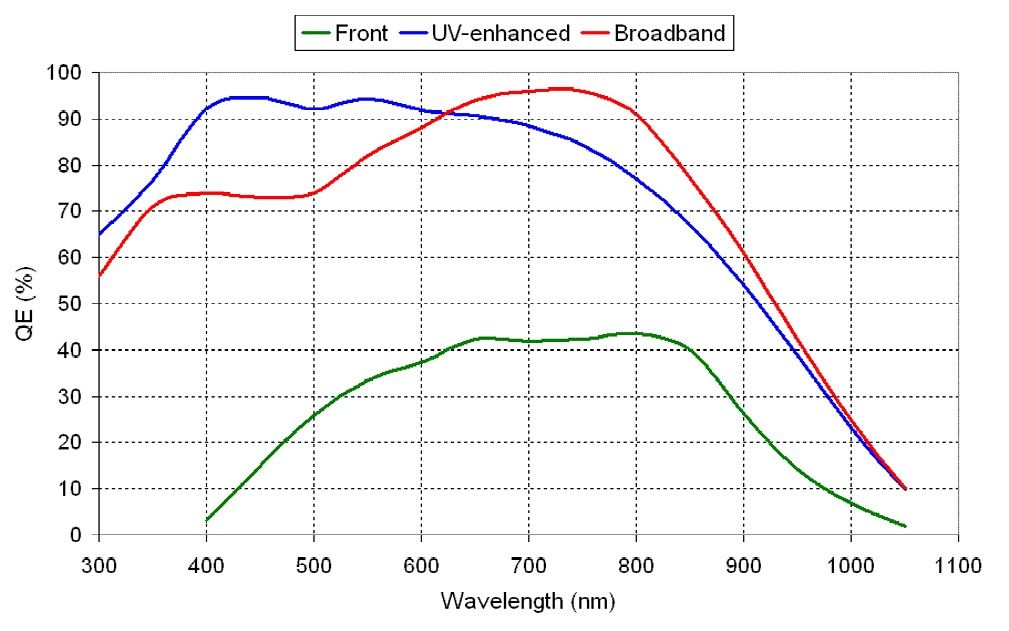}
		\caption{Quantum efficiency (QE) of the Fairchild 486 Back Illuminated chip attached to SI 1100S CCD camera. The UV-enhanced version (solid blue line) of the chip is used in our observations.} 
  \label{fig:fig1}
	\end{center}
\end{figure}

\begin{figure}
	\begin{center}
		\includegraphics[width=0.95\columnwidth]{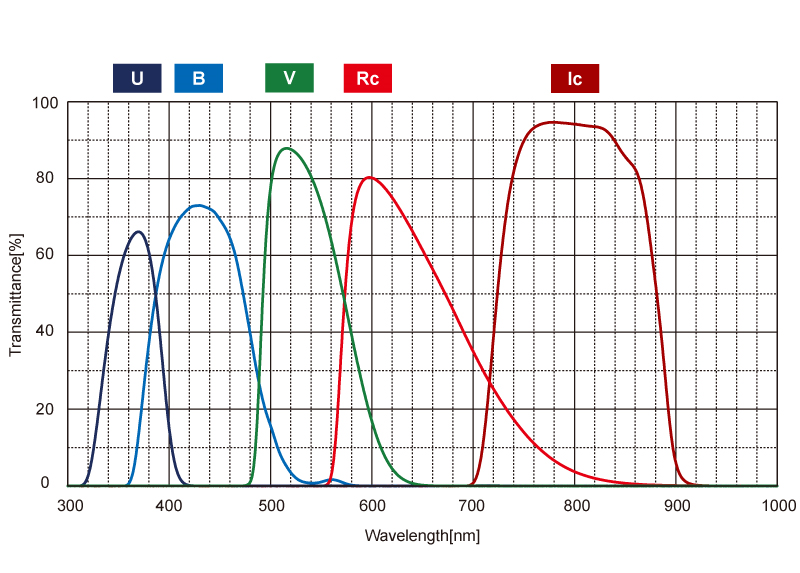}
		\caption{Transmittance curves of the conventional Bessell $UBV$ filters of Asahi.} 
  \label{fig:fig2}
	\end{center}
\end{figure}

All the observations have been performed with the 1-meter telescope (T100) of T\"{U}B\.{I}TAK National Observatory. The T100 telescope has a Ritchey-Chretien optical system with an $f$/10 focal ratio which provides a wide field of view using appropriate 3-element field lenses\footnote{https://tug.tubitak.gov.tr/en}. T100 is equipped with an SI 1100 CCD camera and Bessell $UBV$ filters. Specifications of the camera are given in Table 1$^1$. The camera has a Fairchild 486 Back Illuminated and UV-enhanced chip, which covers a field of view of 21$\arcmin$.5 $\times$ 21$\arcmin$.5. Quantum efficiency (QE) of the chip is shown in Figure \ref{fig:fig1}\footnote{http://linmax.sao.arizona.edu/FLWO/48/CCD486DataSheetRevB.pdf}. QE of the chip across the $U$ passband is very good, i.e. its QE is $\sim$65\% at $\lambda$=300 nm and $\sim$92\% at $\lambda$=400 nm. Transmittance curves of the Bessell $UBV$ filters are presented in Figure \ref{fig:fig2}\footnote{https://www.asahi-spectra.com/}. Note that the $B$ filter has a very weak visual leak centered at $\sim$560.5 nm with a maximum transmittance of $\sim$1.6\%. Since the transmittance of Bessell $U$ filter starts at $\sim$315 nm and peaks at $\sim$370 nm, QE curve of the chip covers this filter with acceptable sensitivity. QE of the chip is also high for the Bessell $B$ and $V$ filters. It should also be noted that all the observations were done with the 2$\times$2 binning mode of the camera to save the data downloading time and observe fainter stars with a high $S/N$ ratio.  

Stellar fields including standard stars selected from \citet{Land09, Land13} have been observed with Bessell $UBV$ filters during 50 nights from 18 July 2012 to 29 September 2024. The fields with red and blue standards were preferred to find the colour dependence of atmospheric extinction.  Each field was observed at least three times using Bessell $UBV$ filters at the same airmass in order to estimate averages of stellar magnitudes. Image Reduction and Analysis Facility (IRAF\footnote{IRAF is distributed by the National Optical Astronomy Observatories}) routines were utilized for pre-reduction processes, bias subtraction and flat fielding the images. We did not perform dark frame subtraction since the camera's dark level is negligible. The instrumental magnitudes of the standard stars were measured utilizing IRAF software packages with aperture photometry. 

%-------------------------------------------------------------------------------------------
%TABLE 1

\begin{table}
\label{tab:tab1}
\setlength{\tabcolsep}{3pt}
\begin{center}
\small{
\caption{Specifications of the SI 1100S camera attached to the T100 telescope of T\"{U}B\.{I}TAK National Observatory.}
\begin{tabular}{ll}
\hline
Camera                 & Spectral Instruments 1100S Cryo, UV, AR, BI   \\
Chip                   & Fairchild 486 Back Illuminated               \\
Read-out channels      & 4 channels                                   \\
Pixel Number	       & 4096 $\times$ 4037                           \\
Pixel Size	           & 15 $\times$ 15 micron                        \\
Chip Size	           & 61.44 $\times$ 61.44 mm                      \\
Gain	               & 0.57 e$^-$/ADU (@ 100 kHz)                   \\
Noise		           & 4.11 e$^-$ (@ 100 kHz)                       \\
Bias level		       & $\sim$500 ADU                                      \\
Dark Current		   & 0.0001  e$^-$/pixel/sec                      \\
Well Depth		       & 142900  e$^-$                                \\
Dynamic Range		   & 16 bit                                       \\
Chip Size	           & 61.44 $\times$ 61.44 mm                      \\
Shutter		           & Bonn 80, Slit Type                           \\
Exposure Range		   & 1 msec to 3600 sec                           \\
Cooling Method		   & Cryo-tiger                                   \\
Operating Temp.        & -100 $^{\circ}$C                             \\
PC Interface	 	   & Gigabit F/O kart (PCI)                       \\
Transfer Time		   & 48 sec (1$\times$1 binning), 13 sec (2$\times$2 binning)   \\
Pixel Scale		       & 0$\arcsec$.31 pixel$^{-1}$                         \\
Field of View		   & $21\arcmin.5 \times 21\arcmin.5$ \\
Software			   & Maxim DL 5.12                                \\
Filter Wheel		   & 2 wheels with 8 holes (76$\times$76 mm each)        \\
Filters			       & Asahi Conventional Bessell $UBVR_{\rm C}I_{\rm C}$       \\
\hline
\end{tabular}  
}
\end{center}
 \end{table}
%---------------------------------------------------------------

%---------------------------------------------------------------
%FIGURE 1
%---------------------------------------------------------------

%---------------------------------------------------------------
%FIGURE 2
%---------------------------------------------------------------

\section{Results}

\subsection{Extinction coefficients and zero points}

Atmospheric extinction is caused primarily by Rayleigh scattering and absorption from gas molecules, dust particles and aerosols in the atmosphere. The amount of extinction depends primarily on airmass but also varies with wavelength and color. These extinction dependencies are corrected by using a primary (or first) extinction coefficient, which depends on airmass, and a secondary extinction coefficient which depends also on colour. In addition, transformation coefficients are needed to transform the extra-atmospheric magnitudes to the standard photometric system. For $V$, $B-V$, and $U-B$, we derived coefficients of the form were given by \citet{Janes2013}
 
 \begin{gather*}
v = V + \alpha_{bv}(B-V) + k_{v}X + C_{bv} \\
b = V + \alpha_{b}(B-V) + k_{b}X + k^{'}_{b}X(B-V) +  C_{b} \\
u = V + (B-V) + \alpha_{ub}(U-B) + k_{u}X + k^{'}_{u}X(U-B) +  C_{ub} \\
\end{gather*}
where $U$, $B$, and $V$ are the magnitudes in the standard photometric system. Parameters $u$, $b$, and $v$ denote the instrumental magnitudes. $X$ is the airmass. Parameters $k$ and $k^{'}$ represent primary and secondary extinction coefficients. $\alpha$ and $C$ are transformation coefficients to the standard photometric system and zero points, respectively. Multiple linear regression fits were applied to the transformation equations given above to estimate the photometric extinction and transformation coefficients with zero points for the observing nights. We determined atmospheric extinction and transformation coefficients under photometric conditions. Since we observed a considerable number of standard stars at different airmasses, we could obtain precise coefficients. The number of usable data points, the atmospheric extinction coefficients and zero points are given in Table~\textcolor{blue}{2}.

%-------------------------------------------------------------------------------------------
%TABLE 3

\begin{table*}
\label{tab:tab3}
\setlength{\tabcolsep}{1.1pt}
\renewcommand{\arraystretch}{1.2}
\begin{center}
%\tiny{
\footnotesize{
\caption{The atmospheric extinction coefficients and zero points obtained each observing night. 
Numbers in parentheses denote the number of usable data points. Median values are given in the 
last line, where errors are standard deviations of the individual values.}
\begin{tabular}{lcccccccc}
\hline
Date       & $k_{\rm u}$         &  $k_{\rm b}$         &  $k_{\rm  v}$        &  $k_{\rm u^{'}}$ &  $k_{\rm b^{'}}$ & $C_{\rm b}$     & $C_{\rm bv}$        &  $C_{\rm ub}$      \\
\hline
2012.07.18 & 0.597$\pm$0.032(38)  & 0.392$\pm$0.022(52)  & 0.247$\pm$0.005(52)  & -0.058$\pm$0.032 & -0.030$\pm$0.019 & 0.771$\pm$0.050 & 0.799$\pm$0.010 & 3.111$\pm$0.054 \\
2012.07.19 & 0.472$\pm$0.031(82)  & 0.326$\pm$0.024(85)  & 0.189$\pm$0.024(85)  & -0.019$\pm$0.032 & -0.057$\pm$0.025 & 0.745$\pm$0.036 & 0.799$\pm$0.017 & 3.143$\pm$0.046 \\
2012.07.20 & 0.559$\pm$0.033(39)  & 0.392$\pm$0.011(30)  & 0.235$\pm$0.002(41)  & -0.002$\pm$0.035 & -0.037$\pm$0.011 & 0.786$\pm$0.017 & 0.838$\pm$0.005 & 3.190$\pm$0.054 \\
2012.08.16 & 0.677$\pm$0.030(63)  & 0.525$\pm$0.050(58)  & 0.242$\pm$0.007(73)  & -0.023$\pm$0.067 & -0.246$\pm$0.079 & 0.786$\pm$0.017 & 0.779$\pm$0.016 & 3.041$\pm$0.043 \\
2012.08.17 & 0.365$\pm$0.025(50)  & 0.270$\pm$0.041(57)  & 0.166$\pm$0.004(66)  & +0.216$\pm$0.041 & +0.002$\pm$0.051 & 0.890$\pm$0.054 & 0.879$\pm$0.015 & 3.453$\pm$0.034 \\
2013.08.08 & 0.397$\pm$0.014(51)  & 0.240$\pm$0.014(72)  & 0.118$\pm$0.003(58)  & -0.007$\pm$0.018 & -0.041$\pm$0.015 & 0.481$\pm$0.021 & 0.538$\pm$0.005 & 2.893$\pm$0.020 \\
2013.08.09 & 0.347$\pm$0.016(99)  & 0.206$\pm$0.014(136) & 0.098$\pm$0.005(124) & -0.066$\pm$0.018 & -0.032$\pm$0.014 & 0.552$\pm$0.022 & 0.591$\pm$0.009 & 2.991$\pm$0.024 \\
2013.08.10 & 0.439$\pm$0.025(79)  & 0.328$\pm$0.042(109) & 0.149$\pm$0.006(116) & -0.126$\pm$0.024 & -0.097$\pm$0.038 & 0.431$\pm$0.060 & 0.559$\pm$0.015 & 2.908$\pm$0.037 \\
2014.08.27 & 0.585$\pm$0.080(41)  & 0.363$\pm$0.076(75)  & 0.273$\pm$0.010(80)  & -0.017$\pm$0.149 & +0.002$\pm$0.097 & 1.124$\pm$0.109 & 1.102$\pm$0.032 & 3.492$\pm$0.111 \\
2014.09.24 & 0.330$\pm$0.071(86)  & 0.263$\pm$0.070(90)  & 0.130$\pm$0.010(102) & -0.038$\pm$0.136 & -0.065$\pm$0.084 & 0.944$\pm$0.104 & 1.006$\pm$0.030 & 3.487$\pm$0.102 \\
2016.08.07 & 0.585$\pm$0.097(74)  & 0.345$\pm$0.053(74)  & 0.176$\pm$0.026(73)  & -0.357$\pm$0.110 & -0.096$\pm$0.056 & 1.238$\pm$0.078 & 1.346$\pm$0.035 & 3.542$\pm$0.133 \\
2016.08.08 & 0.411$\pm$0.066(69)  & 0.243$\pm$0.064(73)  & 0.147$\pm$0.020(72)  & -0.132$\pm$0.073 & -0.017$\pm$0.068 & 1.342$\pm$0.091 & 1.366$\pm$0.030 & 3.778$\pm$0.096 \\
2016.09.28 & 0.311$\pm$0.059(95)  & 0.214$\pm$0.064(96)  & 0.106$\pm$0.024(96)  & -0.025$\pm$0.107 & -0.054$\pm$0.069 & 1.340$\pm$0.090 & 1.381$\pm$0.034 & 3.824$\pm$0.082 \\
2016.10.08 & 0.308$\pm$0.061(80)  & 0.151$\pm$0.041(81)  & 0.104$\pm$0.018(89)  & +0.091$\pm$0.110 & -0.003$\pm$0.048 & 1.444$\pm$0.063 & 1.391$\pm$0.027 & 3.807$\pm$0.089 \\
2018.07.17 & 0.578$\pm$0.128(34)  & 0.351$\pm$0.090(41)  & 0.140$\pm$0.041(49)  & -0.139$\pm$0.146 & -0.108$\pm$0.084 & 1.944$\pm$0.128 & 2.083$\pm$0.060 & 4.216$\pm$0.186 \\
2018.08.13 & 0.403$\pm$0.164(27)  & 0.198$\pm$0.085(31)  & 0.172$\pm$0.048(30)  & +0.072$\pm$0.176 & +0.061$\pm$0.090 & 1.490$\pm$0.126 & 1.449$\pm$0.074 & 3.778$\pm$0.241 \\
2018.08.14 & 0.266$\pm$0.121(61)  & 0.195$\pm$0.056(55)  & 0.103$\pm$0.023(54)  & +0.266$\pm$0.121 & -0.034$\pm$0.056 & 1.476$\pm$0.081 & 1.540$\pm$0.036 & 3.960$\pm$0.170 \\
2018.10.06 & 0.580$\pm$0.062(44)  & 0.262$\pm$0.047(41)  & 0.148$\pm$0.019(44)  & -0.075$\pm$0.079 & -0.030$\pm$0.057 & 1.391$\pm$0.071 & 1.475$\pm$0.031 & 3.538$\pm$0.097 \\
2018.11.05 & 0.502$\pm$0.078(43)  & 0.232$\pm$0.046(46)  & 0.128$\pm$0.020(51)  & -0.117$\pm$0.095 & -0.037$\pm$0.050 & 1.488$\pm$0.070 & 1.550$\pm$0.031 & 3.724$\pm$0.121 \\
2018.11.06 & 0.521$\pm$0.062(44)  & 0.226$\pm$0.046(45)  & 0.124$\pm$0.018(54)  & -0.122$\pm$0.069 & -0.021$\pm$0.052 & 1.512$\pm$0.068 & 1.570$\pm$0.028 & 3.718$\pm$0.093 \\
2019.07.30 & 0.498$\pm$0.092(45)  & 0.293$\pm$0.079(50)  & 0.208$\pm$0.026(49)  & -0.068$\pm$0.108 & +0.026$\pm$0.082 & 1.799$\pm$0.116 & 1.782$\pm$0.037 & 4.162$\pm$0.116 \\
2019.09.29 & 0.471$\pm$0.074(57)  & 0.298$\pm$0.061(72)  & 0.175$\pm$0.021(73)  & +0.101$\pm$0.114 & -0.004$\pm$0.072 & 1.783$\pm$0.091 & 1.834$\pm$0.032 & 4.176$\pm$0.110 \\
2019.09.30 & 0.413$\pm$0.069(66)  & 0.285$\pm$0.061(78)  & 0.182$\pm$0.022(84)  & +0.014$\pm$0.094 & +0.029$\pm$0.069 & 1.850$\pm$0.092 & 1.853$\pm$0.033 & 4.338$\pm$0.102 \\
2020.07.21 & 0.533$\pm$0.124(53)  & 0.307$\pm$0.052(52)  & 0.119$\pm$0.018(47)  & -0.322$\pm$0.098 & -0.106$\pm$0.040 & 2.089$\pm$0.079 & 2.202$\pm$0.025 & 4.421$\pm$0.182 \\
2020.07.22 & 0.465$\pm$0.130(43)  & 0.431$\pm$0.081(45)  & 0.157$\pm$0.032(52)  & -0.221$\pm$0.133 & -0.237$\pm$0.085 & 1.904$\pm$0.124 & 2.151$\pm$0.048 & 4.470$\pm$0.191 \\
2020.07.23 & 0.493$\pm$0.116(42)  & 0.272$\pm$0.058(45)  & 0.135$\pm$0.022(52)  & -0.272$\pm$0.107 & -0.097$\pm$0.050 & 2.130$\pm$0.088 & 2.174$\pm$0.033 & 4.491$\pm$0.169 \\
2021.07.06 & 0.469$\pm$0.155(30)  & 0.273$\pm$0.085(30)  & 0.193$\pm$0.035(39)  & -0.200$\pm$0.117 & +0.003$\pm$0.070 & 2.754$\pm$0.111 & 2.640$\pm$0.044 & 5.178$\pm$0.198 \\
2021.07.07 & 0.370$\pm$0.083(50)  & 0.319$\pm$0.064(56)  & 0.230$\pm$0.025(58)  & -0.093$\pm$0.086 & +0.053$\pm$0.063 & 2.733$\pm$0.111 & 2.624$\pm$0.033 & 5.318$\pm$0.108 \\
2021.10.08 & 0.393$\pm$0.027(65)  & 0.218$\pm$0.018(82)  & 0.154$\pm$0.007(78)  & +0.055$\pm$0.046 & +0.009$\pm$0.063 & 2.796$\pm$0.029 & 2.673$\pm$0.012 & 5.216$\pm$0.042 \\
2021.10.09 & 0.399$\pm$0.029(67)  & 0.244$\pm$0.020(76)  & 0.139$\pm$0.007(74)  & -0.005$\pm$0.047 & -0.004$\pm$0.022 & 2.757$\pm$0.031 & 2.699$\pm$0.012 & 5.195$\pm$0.044 \\
2021.10.11 & 0.607$\pm$0.056(58)  & 0.395$\pm$0.064(70)  & 0.211$\pm$0.019(65)  & +0.414$\pm$0.072 & -0.080$\pm$0.083 & 2.684$\pm$0.079 & 2.710$\pm$0.025 & 5.090$\pm$0.071 \\
2022.06.23 & 0.712$\pm$0.069(33)  & 0.560$\pm$0.093(34)  & 0.305$\pm$0.026(35)  & -0.472$\pm$0.128 & -0.204$\pm$0.124 & 2.948$\pm$0.123 & 3.014$\pm$0.032 & 5.428$\pm$0.092 \\
2022.08.04 & 0.566$\pm$0.154(72)  & 0.354$\pm$0.048(67)  & 0.231$\pm$0.016(60)  & +0.326$\pm$0.165 & -0.058$\pm$0.051 & 3.123$\pm$0.064 & 3.018$\pm$0.023 & 5.500$\pm$0.709 \\
2022.08.31 & 0.470$\pm$0.067(71)  & 0.263$\pm$0.045(70)  & 0.182$\pm$0.014(65)  & -0.005$\pm$0.056 & -0.025$\pm$0.044 & 0.558$\pm$0.058 & 0.605$\pm$0.019 & 2.916$\pm$0.084 \\
2022.09.01 & 0.641$\pm$0.123(72)  & 0.486$\pm$0.124(74)  & 0.173$\pm$0.014(67)  & -0.153$\pm$0.164 & -0.263$\pm$0.151 & 0.342$\pm$0.160 & 0.622$\pm$0.018 & 2.757$\pm$0.159 \\
2022.09.21 & 0.553$\pm$0.060(81)  & 0.330$\pm$0.045(85)  & 0.177$\pm$0.015(92)  & -0.078$\pm$0.071 & -0.050$\pm$0.048 & 0.561$\pm$0.058 & 0.658$\pm$0.020 & 2.924$\pm$0.020 \\
2022.09.22 & 0.536$\pm$0.092(70)  & 0.318$\pm$0.052(79)  & 0.133$\pm$0.018(90)  & -0.354$\pm$0.136 & -0.125$\pm$0.061 & 0.579$\pm$0.068 & 0.710$\pm$0.023 & 2.942$\pm$0.116 \\
2022.10.26 & 0.454$\pm$0.048(95)  & 0.343$\pm$0.039(102) & 0.157$\pm$0.015(113) & -0.224$\pm$0.061 & -0.110$\pm$0.038 & 0.576$\pm$0.048 & 0.718$\pm$0.018 & 3.035$\pm$0.059 \\
2022.10.27 & 0.410$\pm$0.044(97)  & 0.255$\pm$0.030(99)  & 0.167$\pm$0.014(104) & -0.068$\pm$0.059 & +0.005$\pm$0.034 & 0.670$\pm$0.040 & 0.695$\pm$0.018 & 3.101$\pm$0.057 \\
2022.12.21 & 0.482$\pm$0.025(86)  & 0.235$\pm$0.024(79)  & 0.170$\pm$0.008(84)  & -0.087$\pm$0.038 & +0.002$\pm$0.028 & 0.780$\pm$0.033 & 0.729$\pm$0.013 & 3.121$\pm$0.036 \\
2023.01.19 & 0.383$\pm$0.037(87)  & 0.275$\pm$0.039(85)  & 0.128$\pm$0.010(83)  & -0.024$\pm$0.045 & -0.068$\pm$0.043 & 0.729$\pm$0.050 & 0.779$\pm$0.013 & 3.266$\pm$0.048 \\
2023.08.16 & 0.468$\pm$0.054(99)  & 0.413$\pm$0.043(95)  & 0.270$\pm$0.019(97)  & +0.087$\pm$0.064 & -0.100$\pm$0.050 & 0.853$\pm$0.054 & 0.950$\pm$0.023 & 3.442$\pm$0.067 \\
2023.08.22 & 0.474$\pm$0.057(103) & 0.312$\pm$0.046(111) & 0.214$\pm$0.020(115) & +0.015$\pm$0.073 & +0.019$\pm$0.050 & 0.970$\pm$0.056 & 0.970$\pm$0.024 & 3.414$\pm$0.070 \\
2024.04.28 & 0.545$\pm$0.062(83)  & 0.445$\pm$0.069(84)  & 0.230$\pm$0.021(87)  & -0.175$\pm$0.092 & -0.156$\pm$0.082 & 1.049$\pm$0.088 & 1.187$\pm$0.027 & 3.582$\pm$0.076 \\
2024.06.10 & 0.530$\pm$0.051(73)  & 0.334$\pm$0.054(80)  & 0.205$\pm$0.016(81)  & -0.133$\pm$0.080 & -0.024$\pm$0.060 & 1.218$\pm$0.072 & 1.263$\pm$0.022 & 3.608$\pm$0.065 \\
2024.06.11 & 0.503$\pm$0.058(67)  & 0.418$\pm$0.062(64)  & 0.277$\pm$0.018(72)  & +0.028$\pm$0.075 & -0.025$\pm$0.067 & 1.178$\pm$0.082 & 1.232$\pm$0.024 & 3.707$\pm$0.073 \\
2024.07.09 & 0.481$\pm$0.061(52)  & 0.290$\pm$0.060(68)  & 0.208$\pm$0.021(72)  & -0.177$\pm$0.152 & +0.026$\pm$0.078 & 1.409$\pm$0.082 & 1.351$\pm$0.028 & 3.867$\pm$0.078 \\
2024.08.01 & 0.561$\pm$0.059(76)  & 0.385$\pm$0.063(74)  & 0.204$\pm$0.020(77)  & -0.006$\pm$0.073 & -0.081$\pm$0.064 & 1.310$\pm$0.077 & 1.368$\pm$0.025 & 3.767$\pm$0.076 \\
2024.08.09 & 0.515$\pm$0.046(74)  & 0.330$\pm$0.049(80)  & 0.195$\pm$0.015(79)  & +0.073$\pm$0.061 & -0.034$\pm$0.052 & 1.428$\pm$0.064 & 1.407$\pm$0.021 & 3.897$\pm$0.062 \\
2024.09.29 & 0.522$\pm$0.038(100) & 0.279$\pm$0.043(99)  & 0.177$\pm$0.016(102) & +0.061$\pm$0.055 & -0.017$\pm$0.052 & 0.926$\pm$0.055 & 0.976$\pm$0.023 & 3.236$\pm$0.050 \\
\hline 
 $\bf{Median}$ & $\bf{0.481\pm0.097}$ & $\bf{0.303\pm0.086}$ & $\bf{0.174\pm0.050}$ & $\bf{-0.048\pm0.164}$ & $\bf{-0.034\pm0.072}$ & $\bf{1.274\pm0.751}$ & $\bf{1.384\pm0.713}$ & $\bf{3.658\pm0.755}$ \\
\hline
\end{tabular}  
}
\end{center}
 \end{table*}
%---------------------------------------------------------------

\begin{figure}
	\begin{center}
		\includegraphics[width=0.50\columnwidth]{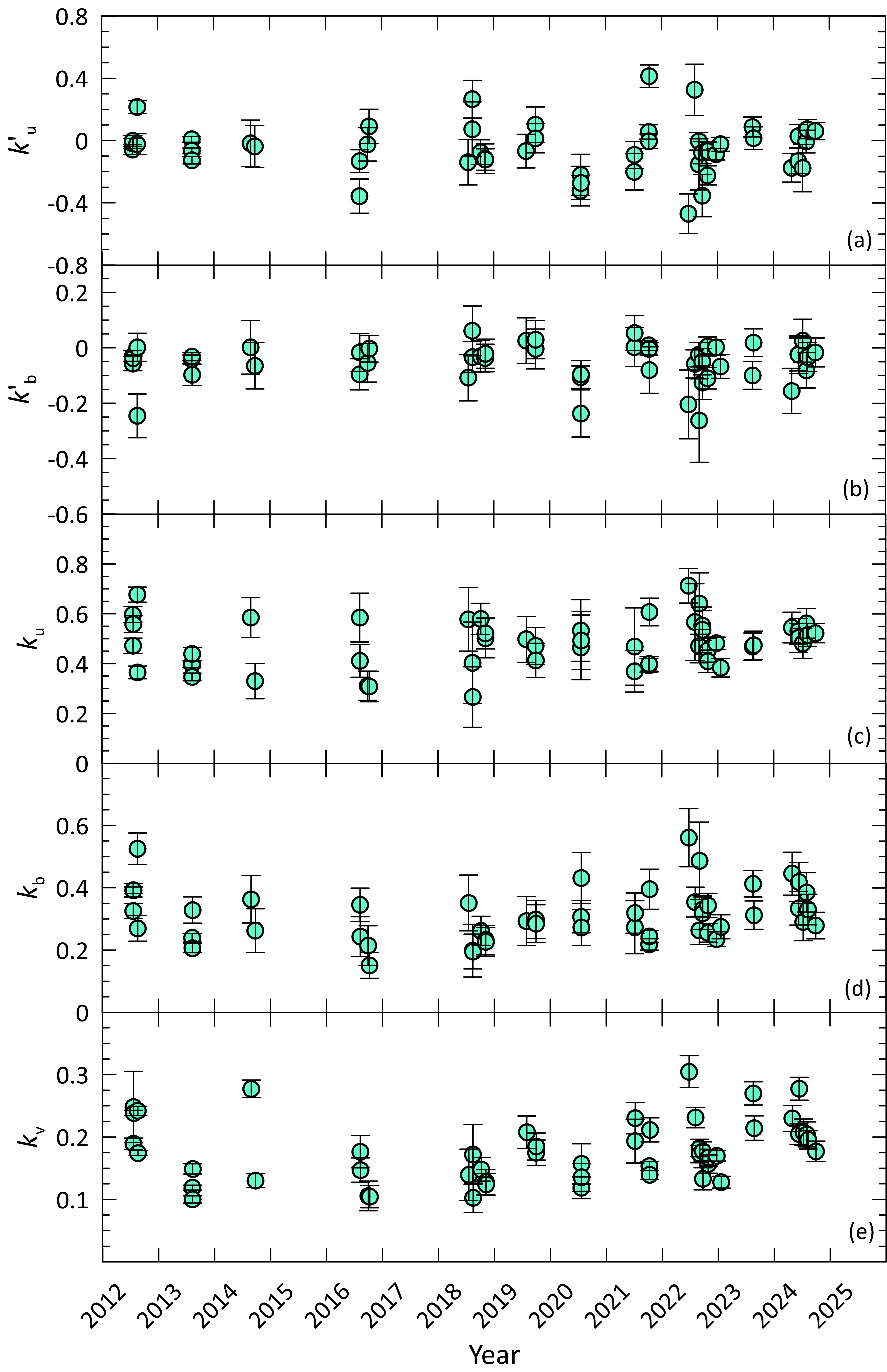}
		\caption{Variation of the primary and secondary extinction coefficients from 2012 to 2024.} 
  \label{fig:fig3}
	\end{center}
\end{figure}

The extinction coefficients in Table 2 span a 12-year observing period, although no ebservations were conducted in 2015 and 2017. The extinction 
coefficients in Table~\textcolor{blue}{2} cover a 12-year observing time, although no observations were conducted in 2015 and 2017. Median values of $k_{\rm u}$, $k_{\rm b}$ and $k_{\rm  v}$ were calculated as $0.481\pm0.097$, $0.303\pm0.086$ and $0.174\pm0.050$, respectively. Median secondary extinction coefficients $k^{'}_{u}$ and $k^{'}_{b}$ were found to be $-0.048\pm0.164$ and $-0.034\pm0.072$. Error-values are standard deviations of the coefficients. Primary and secondary extinction coefficients can vary during the years depending on the atmospheric conditions of the observatory. Figure \ref{fig:fig3} exhibits such a slight variation of $k_{\rm u}$, $k_{\rm b}$ and $k_{\rm  v}$, where the increase after the year 2019 is prominent for $V$ and $B$ bands. 
These increases in extinction coefficients suggest that the photometric conditions at the observatory have gradually deteriorated since 2019. We could not detect considerable systematic increase or decrease in secondary extinction coefficients $k^{'}_{u}$ and $k^{'}_{b}$. 

Seasonal variations in primary extinction coefficients can help select appropriate observing nights for research projects. Unfortunately, it is not possible to find reliable median values of extinction coefficients obtained between December and February, since we have standard star observations for only two nights in this interval during the observing period of 12 years. Mean values of extinction coefficients of these two winter nights are  $k_{\rm u}$=0.432$\pm$0.031, $k_{\rm b}$=0.255$\pm$0.031 and $k_{\rm v}$=0.149$\pm$0.009, where errors are mean values of individual errors. Similarly, we could observe standard stars only one night between March and May during 12 years of observing period. The remaining observations were performed summer (June-August) and autumn (September-November) seasons. Median values of extinction coefficients are $k_{\rm u}$=0.477$\pm$0.089, $k_{\rm b}$=0.322$\pm$0.077 and $k_{\rm v}$=0.191$\pm$0.051 for summer season, while $k_{\rm u}$=0.502$\pm$0.113, $k_{\rm b}$=0.279$\pm$0.098 and $k_{\rm v}$=0.157$\pm$0.047 for autumn season. There appears to be no significant seasonal difference in extinction coefficients between summer and autumn, as their median values are very similar within the margins of error. Based on the seasonal extinction coefficients and the number of usable nights, it is evident that winter and spring are not favorable seasons for photometric observations at the T\"{U}B\.{I}TAK National Observatory.

\begin{figure}
	\begin{center}
		\includegraphics[width=0.50\columnwidth]{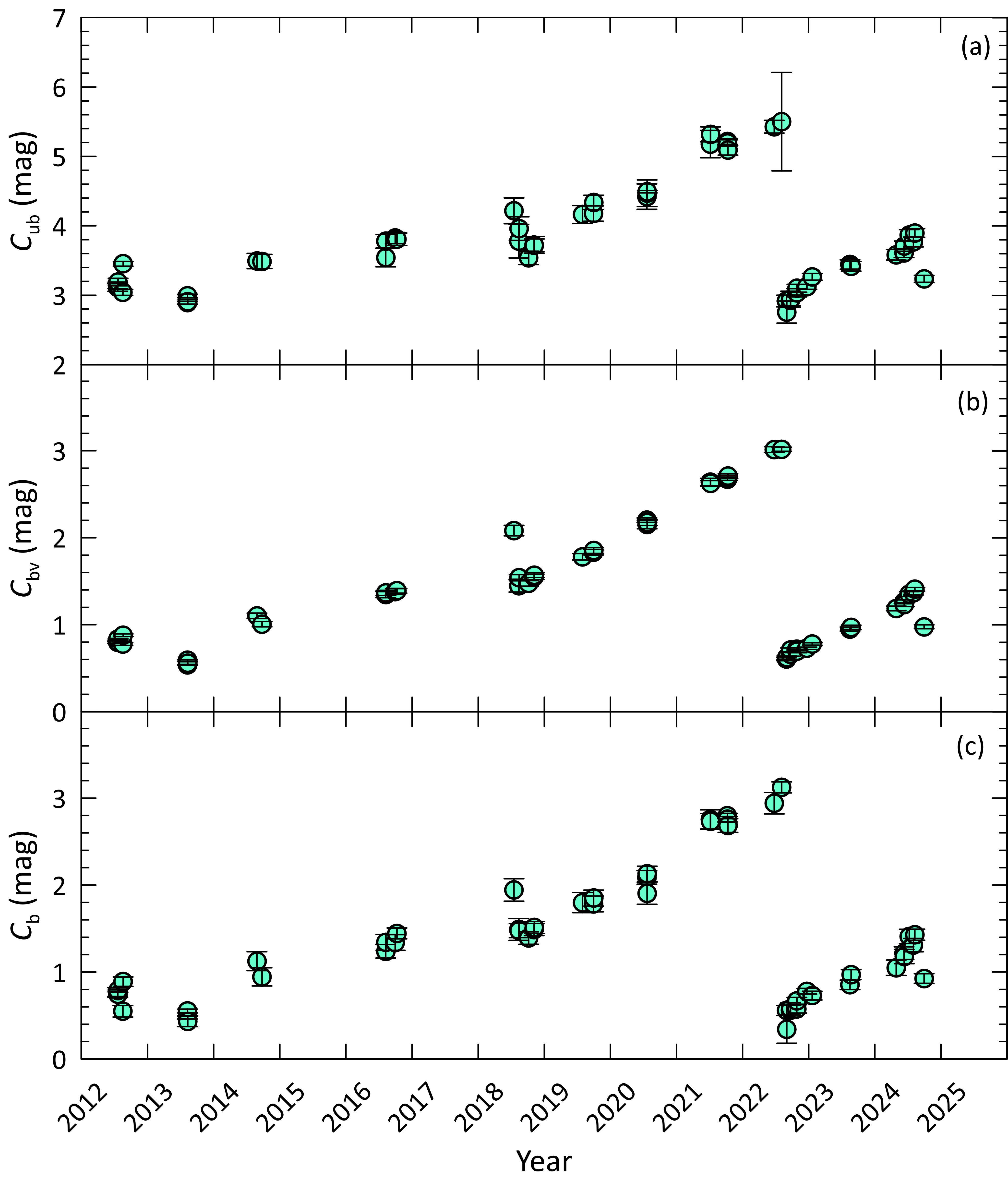}
		\caption{Variation of zero points from 2012 to 2024.} 
  \label{fig:fig4}
	\end{center}
\end{figure}

It is known that instrumental parameters and atmospheric conditions affect the photometric zero point. The value of the photometric zero point depends on the size and condition (primarily mirror reflectivity) of the telescope and the quantum efficiency of the detector. Atmospheric conditions, such as water vapour content and height of the ozone layer, also affect the photometric zero points. The photometric zero points $C_{\rm b}$, $C_{\rm bv}$ and $C_{\rm ub}$ measured during our observations are listed in Table~\textcolor{blue}{2}. Variations of zero points are shown in Figure \ref{fig:fig4}. As can be seen in Figure \ref{fig:fig4}, variation of the zero points with time clearly exhibits the condition of the telescopic reflectivity. Beginning with the year 2012, reflectivity decreases (zero points become fainter) with time. Cleaning of the main mirror in August 2022 can be seen in Figure \ref{fig:fig4} as a sudden brightening of zero points.

%---------------------------------------------------------------
%FIGURE 3
%---------------------------------------------------------------

%---------------------------------------------------------------
%FIGURE 4
%---------------------------------------------------------------

%-------------------------------------------------------------------------------------------
%TABLE 2

\begin{table}
\label{tab:tab2}
\setlength{\tabcolsep}{5pt}
\renewcommand{\arraystretch}{1.2}
\begin{center}
\small{
\caption{The transformation coefficients calculated for each observing night. Median values are given in the last line, where errors are standard deviations of the individual values.}
\begin{tabular}{lccc}
\hline
Date       & $\alpha_{b}$     &  $\alpha_{bv}$   &  $\alpha_{ub}$   \\
\hline
2012.07.18 & 0.930$\pm$0.043  & 0.056$\pm$0.006  & 0.926$\pm$0.054   \\
2012.07.19 & 0.992$\pm$0.038  & 0.077$\pm$0.011  & 0.861$\pm$0.050   \\
2012.07.20 & 0.936$\pm$0.018  & 0.051$\pm$0.003  & 0.813$\pm$0.056   \\
2012.08.16 & 1.253$\pm$0.105  & 0.084$\pm$0.011  & 0.836$\pm$0.093   \\
2012.08.17 & 0.919$\pm$0.066  & 0.081$\pm$0.011  & 0.537$\pm$0.055   \\
2013.08.08 & 0.972$\pm$0.024  & 0.070$\pm$0.003  & 0.840$\pm$0.027   \\
2013.08.09 & 0.959$\pm$0.023  & 0.068$\pm$0.006  & 0.951$\pm$0.028  \\
2013.08.10 & 1.045$\pm$0.055  & 0.069$\pm$0.010  & 1.038$\pm$0.037 \\
2014.08.27 & 0.869$\pm$0.137  & 0.041$\pm$0.022  & 0.853$\pm$0.221   \\
2014.09.24 & 1.004$\pm$0.124  & 0.086$\pm$0.020  & 0.929$\pm$0.196  \\
2016.08.07 & 1.031$\pm$0.086  & 0.057$\pm$0.009  & 1.378$\pm$0.166   \\
2016.08.08 & 0.920$\pm$0.095  & 0.056$\pm$0.007  & 0.613$\pm$0.105   \\
2016.09.28 & 0.968$\pm$0.099  & 0.058$\pm$0.010  & 0.849$\pm$0.146)   \\
2016.10.08 & 0.900$\pm$0.073  & 0.068$\pm$0.008  & 0.754$\pm$0.158   \\
2018.07.17 & 1.046$\pm$0.122  & 0.049$\pm$0.014  & 1.023$\pm$0.218  \\
2018.08.13 & 0.805$\pm$0.135  & 0.045$\pm$0.017  & 0.730$\pm$0.266  \\
2018.08.14 & 0.967$\pm$0.079  & 0.072$\pm$0.008  & 0.444$\pm$0.172  \\
2018.10.06 & 0.978$\pm$0.084  & 0.101$\pm$0.011  & 0.988$\pm$0.120   \\
2018.11.05 & 0.968$\pm$0.076  & 0.078$\pm$0.008  & 0.977$\pm$0.142   \\
2018.11.06 & 0.932$\pm$0.077  & 0.073$\pm$0.007  & 0.986$\pm$0.102   \\
2019.07.30 & 0.866$\pm$0.127  & 0.073$\pm$0.011  & 0.953$\pm$0.161  \\
2019.09.29 & 0.900$\pm$0.107  & 0.062$\pm$0.008  & 0.684$\pm$0.166   \\
2019.09.30 & 0.852$\pm$0.105  & 0.066$\pm$0.008  & 0.801$\pm$0.137   \\
2020.07.21 & 1.057$\pm$0.064  & 0.071$\pm$0.009  & 1.304$\pm$0.157   \\
2020.07.22 & 1.245$\pm$0.132  & 0.073$\pm$0.014  & 1.203$\pm$0.203   \\
2020.07.23 & 1.044$\pm$0.081  & 0.066$\pm$0.010  & 1.221$\pm$0.161   \\
2021.07.06 & 0.874$\pm$0.099  & 0.060$\pm$0.014  & 1.043$\pm$0.163   \\
2021.07.07 & 0.810$\pm$0.090  & 0.063$\pm$0.013  & 0.882$\pm$0.115   \\
2021.10.08 & 0.893$\pm$0.030  & 0.088$\pm$0.007  & 0.776$\pm$0.066  \\
2021.10.09 & 0.920$\pm$0.035  & 0.078$\pm$0.006  & 0.891$\pm$0.071   \\
2021.10.11 & 0.988$\pm$0.101  & 0.053$\pm$0.010  & 0.282$\pm$0.094   \\
2022.06.23 & 1.165$\pm$0.163  & 0.082$\pm$0.013  & 1.426$\pm$0.166   \\
2022.08.04 & 0.961$\pm$0.070  & 0.074$\pm$0.010  & 0.565$\pm$0.900   \\
2022.08.31 & 0.956$\pm$0.059  & 0.065$\pm$0.008  & 0.842$\pm$0.074   \\
2022.09.01 & 1.222$\pm$0.194  & 0.084$\pm$0.008  & 0.999$\pm$0.213   \\
2022.09.21 & 0.968$\pm$0.063  & 0.074$\pm$0.008  & 0.945$\pm$0.096   \\
2022.09.22 & 1.092$\pm$0.084  & 0.091$\pm$0.010  & 1.422$\pm$0.179  \\
2022.10.26 & 1.049$\pm$0.049  & 0.071$\pm$0.007  & 1.169$\pm$0.084  \\
2022.10.27 & 0.920$\pm$0.046  & 0.081$\pm$0.008  & 0.972$\pm$0.085  \\
2022.12.21 & 0.888$\pm$0.039  & 0.079$\pm$0.008  & 1.010$\pm$0.058   \\
2023.01.19 & 0.958$\pm$0.058  & 0.065$\pm$0.008  & 0.855$\pm$0.064   \\
2023.08.16 & 1.054$\pm$0.067  & 0.050$\pm$0.023  & 0.715$\pm$0.082   \\
2023.08.22 & 0.890$\pm$0.064  & 0.076$\pm$0.009  & 0.840$\pm$0.093  \\
2024.04.28 & 1.075$\pm$0.106  & 0.046$\pm$0.013  & 1.038$\pm$0.116   \\
2024.06.10 & 0.928$\pm$0.082  & 0.060$\pm$0.012  & 1.027$\pm$0.106   \\
2024.06.11 & 0.926$\pm$0.093  & 0.067$\pm$0.012  & 0.797$\pm$0.096  \\
2024.07.09 & 0.861$\pm$0.112  & 0.071$\pm$0.012  & 0.956$\pm$0.175  \\
2024.08.01 & 1.001$\pm$0.081  & 0.071$\pm$0.010  & 0.835$\pm$0.094  \\
2024.08.09 & 0.947$\pm$0.069  & 0.082$\pm$0.010  & 0.681$\pm$0.083   \\
2024.09.29 & 0.926$\pm$0.068  & 0.062$\pm$0.011  & 0.738$\pm$0.073  \\
\hline
$\bf{Median}$ & $\bf{0.958\pm0.100}$ & $\bf{0.070\pm0.013}$ & $\bf{0.886\pm0.228}$  \\
\hline
\end{tabular}  
}
\end{center}
 \end{table}
%---------------------------------------------------------------

\subsection{Transformation coefficients}
Transformation coefficients found from the standard star observations made during 50 nights between the years 2012 and 2024 are given in Table~\textcolor{blue}{3}. Median values of the transformation coefficients are $\alpha_{\rm b}$=$0.958\pm0.100$, $\alpha_{\rm bv}$=$0.070\pm0.013$ and $\alpha_{\rm ub}$=$0.886\pm0.228$, where errors are standard deviations of the individual values. We calculated possible maximum values by adding standard deviations to the median transformation coefficients. Using these maximum values, the resulting magnitudes differ by at most 25-30 mmag for a red star (B - V = 1.9mag) compared to those calculated with median coefficients, assuming extinction coefficients and zero points remain constant.

\subsection{Sources of extinction}

\begin{figure*}[h]
	\begin{center}
		\includegraphics[width=0.80\columnwidth]{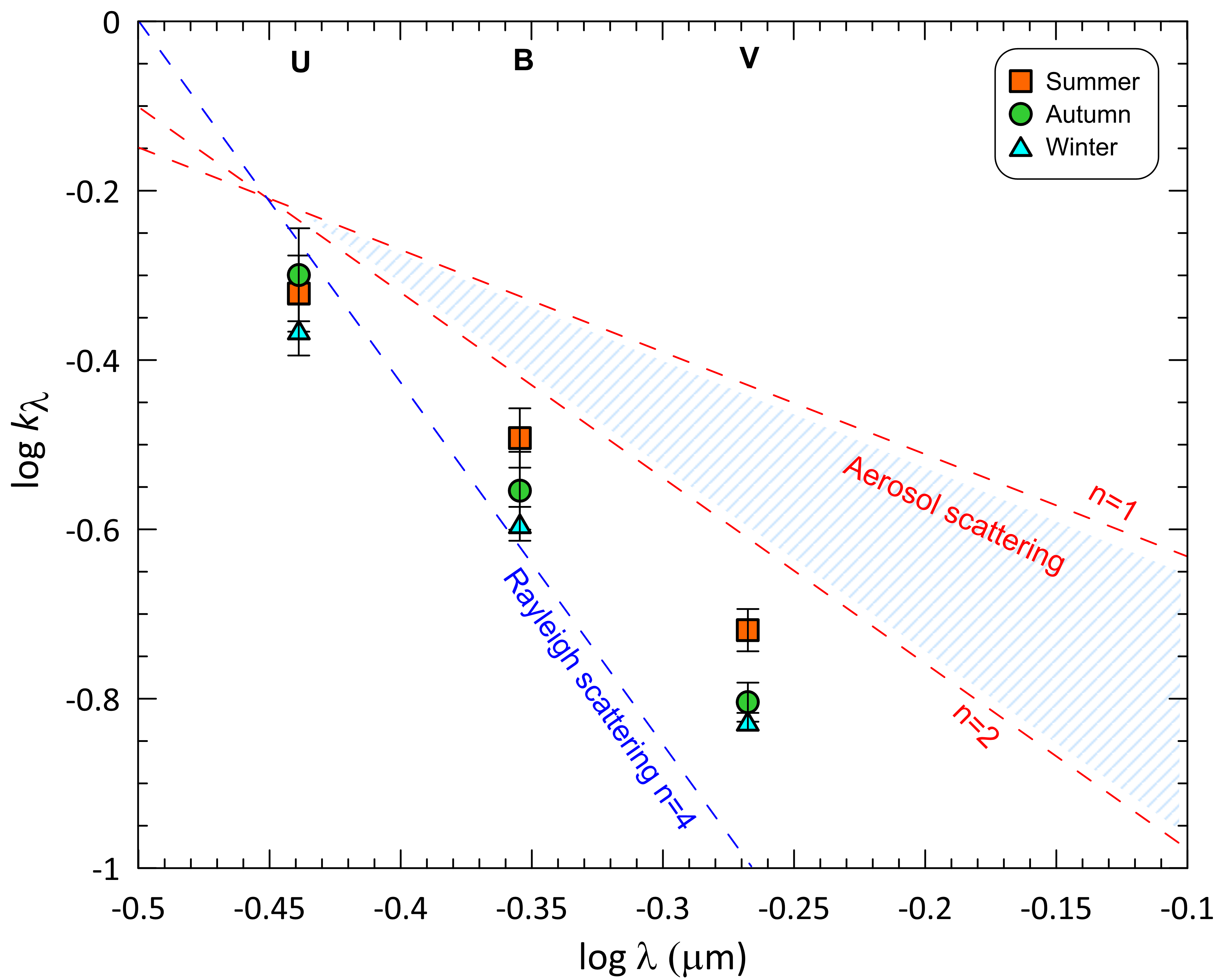}
		\caption{Variation of seasonal median values of extinction coefficients with wavelength. The shaded part represents the area affected by scattering due to aerosols and dust, while the line with $n$=4 represents the pure Rayleigh scattering.} 
  \label{fig:fig5}
	\end{center}
\end{figure*}

Small seasonal differences in median extinction coefficients may result from extinction sources in the atmosphere. Atmospheric extinction is mainly a result of scattered light from molecules and small particles. The scattering efficiency depends on wavelength. The relation between the atmospheric extinction coefficients and wavelength can be expressed as $k_{\lambda}=\beta/\lambda^{n}$, where $k_{\lambda}$, $\beta$ and $\lambda$ are the extinction coefficient, an appropriate constant and the mean wavelength of filter \citep{Golay1974}. If extinction is due to the Rayleigh scattering, then $n$=4. When extinction is caused by aerosol and dust, then $n$ is between 1 and 2. Seasonal averages are shown in Figure \ref{fig:fig5}, where extinction coefficient variation with wavelength are also drawn for $n$=1, $n$=2 and $n$=4. Figure \ref{fig:fig5} shows that extinction during winter and autumn is almost entirely due to Rayleigh scattering. For the summer season, the source of extinction is mainly Rayleigh scattering although aerosol scattering has some effect.

%---------------------------------------------------------------
%FIGURE 5
%---------------------------------------------------------------

%---------------------------------------------------------------
%FIGURE 6
\begin{figure*}[h]
	\begin{center}
		\includegraphics[width=0.7\columnwidth]{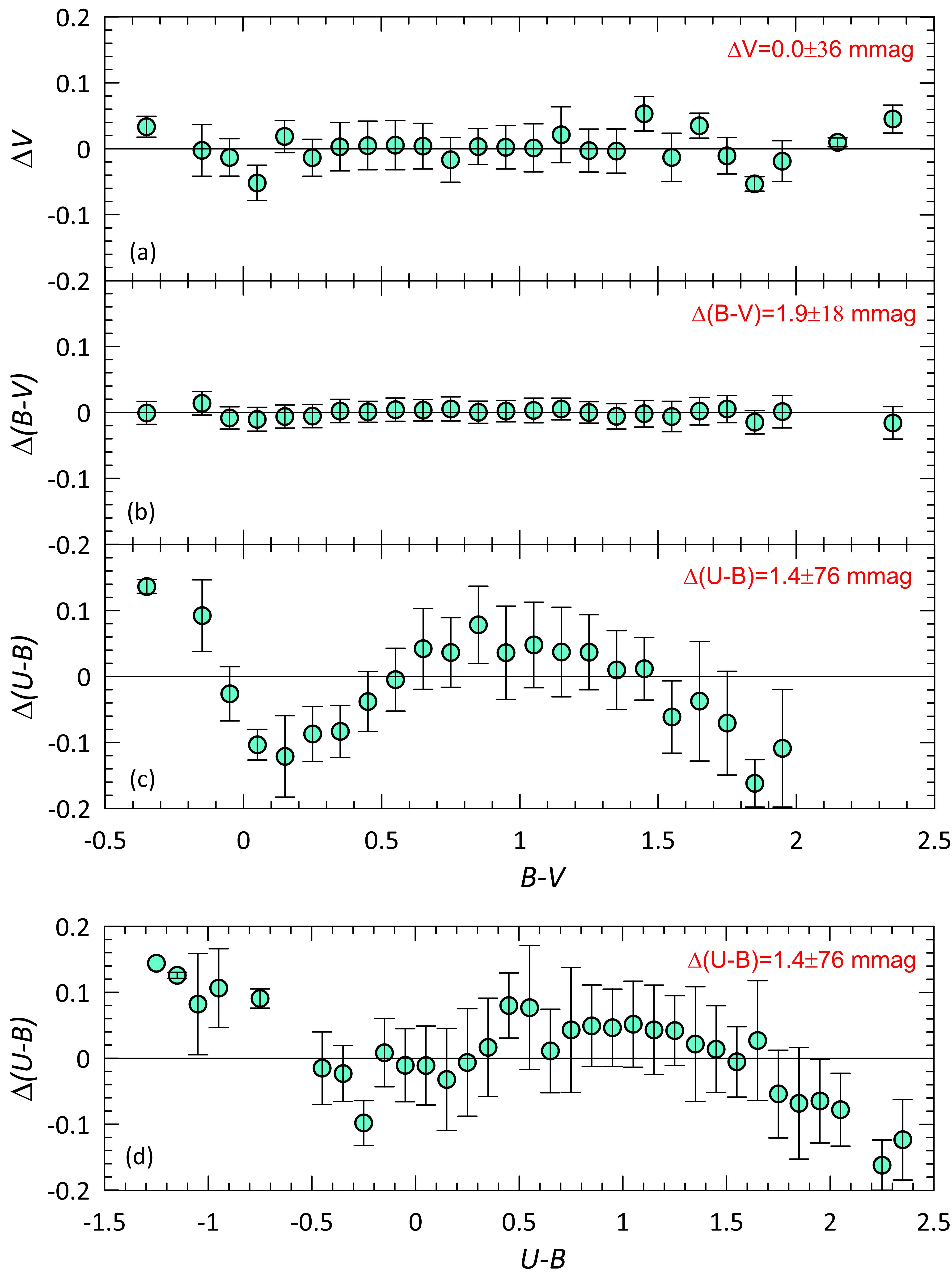}
		\caption{Median $\Delta(U-B)$, $\Delta(B-V)$, and $\Delta V$ values againts colour indices. $\Delta$ means calculated minus catalogue \citep{Land09,Land13} value. Median values were calculated for 0.1 mag intervals of related colour index. Means and standard deviations of $\Delta(U-B)$, $\Delta(B-V)$, and $\Delta V$ are given in panels.} 
  \label{fig:fig6}
	\end{center}
\end{figure*}
%---------------------------------------------------------------
\vspace{-20pt}

\subsection{Comparison with Landolt's catalogue}
In order to find differences between the T100 and Landolt's photometric systems, we estimated differences of standard star' $V$ magnitudes and $U-B$ and $B-V$ colour indices calculated from our transformation equations and the ones taken from Landolt's catalogues \citep{Land09, Land13} for 34 nights between 2018 and 2024. Since there are 2188, 2421 and 2324 standard star observations for $U-B$, $B-V$ and $V$ in these nights, respectively, we calculated median values of the differences for 0.1 mag intervals of $U-B$ and $B-V$ colour indices. The distribution of the median values with respect to the corresponding color indices is shown in Figure \ref{fig:fig6}. Here $\Delta$ indicates the calculated value minus the catalogue value. Means and standard deviations of $\Delta(U-B)$, $\Delta(B-V)$ and $\Delta V$ were estimated to be 1.4$\pm$76, 1.9$\pm$18 and 0.0$\pm$36 mmag, respectively. 

Figure \ref{fig:fig6} reveals that there are systematic differences between the T100's and Landolt's photometric systems for the $U$-band. Differences between the two systems in $U-B$ follow a sinusoidal-like curve against $U-B$ and $B-V$. However, the median differences are relatively small ($|\Delta(U-B)|\leq 0.05$) for stars with $-0.5<U-B~{\rm (mag)}<1.6$ and $0.2<B-V~{\rm (mag)}<1.8$, although it is considerably high for bluer and redder stars. This difference probably originates from the quantum efficiency of detectors used in the two photometric systems. As for the differences of $\Delta(B-V)$ and $\Delta V$, Figure \ref{fig:fig6} shows that transformation equations found for the T100 photometric system work well. The median values of $|\Delta(B-V)|$ are smaller than 15 mmag for stars with $-0.4<B-V~{\rm (mag)}<2.4$, while almost all median values of $|\Delta V|$ are smaller than 20 mmag for stars with $-0.2<B-V~{\rm (mag)}<1.8$. Thus, we conclude that the T100's photometric system well matches that of Landolt's photometric systems for $B-V$ and $V$.

\section{Summary}

We observed many standard stars selected from \citet{Land09, Land13} with the Bessell $UBV$ filters during 50 nights from the year 2012 to 2024 with the 1-meter telescope (T100) of the T\"{U}B\.{I}TAK National Observatory to perform photometric analysis of open clusters. As a byproduct, we derived precise transformation relations for the T100 photometric system. 
 
\begin{enumerate}
  \item[\textbf{1.}] Primary and secondary atmospheric extinction coefficients were determined for nights with photometric conditions. Median values of primary extinction coefficients were found to be $0.481\pm0.097$, $0.303\pm0.086$ and $0.174\pm0.050$ for $U$, $B$ and $V$ filters, respectively. Median secondary extinction coefficients $k^{'}_{u}$ and $k^{'}_{b}$ were calculated as $-0.048\pm0.164$ and $-0.034\pm0.072$, respectively. We found that primary extinction coefficients do not show a strong seasonal variation. We conclude that the median values of extinction coefficients estimated for summer and autumn are very similar within errors. Seasonal values of the coefficients and number of usable nights show that the winter and spring can not be favourite seasons for photometric observations at the T\"{U}B\.{I}TAK National Observatory.
   
   \item[\textbf{2.}] 
   
   Our observations span a 12-year period form 2012 to 2024, excluding the years 2015 and 2017, allowing us to determine the variation in extinction coefficients over this time. We found that primary extinction coefficients decreased from the year 2012 to 2019, while they increased from 2019 to 2024, indicating deterioration of photometric conditions starting from the year 2019. No systematic variation in the secondary extinction coefficients could be identified. 

   \item[\textbf{3.}] The values of photometric zero points for $B$, $B-V$, and $U-B$ gradually become fainter during years, as expected. In addition, we found a "jump" of zero points to brighter magnitudes in August 2022, corresponding to the cleaning of the main mirror of the telescope. 

   \item[\textbf{4.}] We investigated the characteristics of atmospheric extinction based on scattering mechanisms. It is found that Rayleigh scattering is the main reason for atmospheric extinction in autumn and winter seasons, while aerosol scattering has some effect on the extinction in summer. 

    \item[\textbf{5.}] It is found that there are systematic differences for the $U$-band between the T100' and Landolt's photometric systems, although the median differences are relatively small for stars with $-0.5<U-B~{\rm (mag)}<1.6$ and $0.2<B-V~{\rm (mag)}<1.8$. This difference probably originates from the quantum efficiency of detectors used in the two photometric systems. We conclude that transformation equations found for the T100' photometric system work well for $V$ and $B-V$ as the median values of $|\Delta(B-V)|$ and $|\Delta V|$ are small for a wide range of $B-V$ colour index. As a result, we also conclude that the T100's photometric system acceptably well matches that of Landolt's photometric systems for $U-B$, $B-V$, and $V$. 
  
  \item[\textbf{6.}] As a general result, we finally conclude that the transformation relations found in this study can be used for standardized photometry with T100's photometric system.  
 \end{enumerate}

\medskip

\begin{description}
  \item[Peer Review:] Externally peer-reviewed.
  \item[Author Contribution:] Conception/Design of study - T.A., R.C., T.Y.; Data Acquisition - T.A., R.C., T.Y.; Data Analysis/Interpretation - T.A., R.C., T.Y.; ~Drafting Manuscript - T.A., R.C., T.Y.;~Critical Revision of Manuscript - T.A., R.C., T.Y.;~Final Approval and Accountability - T.A., R.C., T.Y.
  \item[Conflict of Interest:] Authors declared no conflict of interest.
  \item[Financial Disclosure:] This study has been supported in part by the Scientific and Technological Research Council (TÜBİTAK) 113F270.
\end{description}

\section*{Acknowledgements}
We thank to TÜBİTAK National Observatory for partial support in using the T100 telescope with project numbers 15AT100-738 and 18CT100-1396. We also thank to the on-duty observers and members of the technical staff at the T\"{U}B\.{I}TAK National Observatory for their support before and during the observations. 

\bibliographystyle{mnras}
\bibliography{Ak}

\begin{thebibliography}{}
\makeatletter
\relax
\def\mn@urlcharsother{\let\do\@makeother \do\$\do\&\do\#\do\^\do\_\do\%\do\~}
\def\mn@doi{\begingroup\mn@urlcharsother \@ifnextchar [ {\mn@doi@} {\mn@doi@[]}}
\def\mn@doi@[#1]#2{\def\@tempa{#1}\ifx\@tempa\@empty \href {http://dx.doi.org/#2} {doi:#2}\else \href {http://dx.doi.org/#2} {#1}\fi \endgroup}
\def\mn@eprint#1#2{\mn@eprint@#1:#2::\@nil}
\def\mn@eprint@arXiv#1{\href {http://arxiv.org/abs/#1} {{\tt arXiv:#1}}}
\def\mn@eprint@dblp#1{\href {http://dblp.uni-trier.de/rec/bibtex/#1.xml} {dblp:#1}}
\def\mn@eprint@#1:#2:#3:#4\@nil{\def\@tempa {#1}\def\@tempb {#2}\def\@tempc {#3}\ifx \@tempc \@empty \let \@tempc \@tempb \let \@tempb \@tempa \fi \ifx \@tempb \@empty \def\@tempb {arXiv}\fi \@ifundefined {mn@eprint@\@tempb}{\@tempb:\@tempc}{\expandafter \expandafter \csname mn@eprint@\@tempb\endcsname \expandafter{\@tempc}}}

\bibitem[\protect\citeauthoryear{{Golay}}{{Golay}}{1974}]{Golay1974}
{Golay} M.,  1974, {Introduction to astronomical photometry}, \mn@doi{10.1007/978-94-010-2169-2.
}

\bibitem[\protect\citeauthoryear{{Janes}, {Barnes}, {Meibom}  \& {Hoq}}{{Janes} et~al.}{2013}]{Janes2013}
{Janes} K.,  {Barnes} S.~A.,  {Meibom} S.,   {Hoq} S.,  2013, \mn@doi [\aj] {10.1088/0004-6256/145/1/7}, \href {https://ui.adsabs.harvard.edu/abs/2013AJ....145....7J} {145, 7}

\bibitem[\protect\citeauthoryear{{Johnson} \& {Morgan}}{{Johnson} \& {Morgan}}{1953}]{Johnson1953}
{Johnson} H.~L.,  {Morgan} W.~W.,  1953, \mn@doi [\apj] {10.1086/145697}, \href {https://ui.adsabs.harvard.edu/abs/1953ApJ...117..313J} {117, 313}

\bibitem[\protect\citeauthoryear{{Kilkenny}, {van Wyk}, {Roberts}, {Marang}  \& {Cooper}}{{Kilkenny} et~al.}{1998}]{Kilkenny1998}
{Kilkenny} D.,  {van Wyk} F.,  {Roberts} G.,  {Marang} F.,   {Cooper} D.,  1998, \mn@doi [\mnras] {10.1046/j.1365-8711.1998.01222.x}, \href {https://ui.adsabs.harvard.edu/abs/1998MNRAS.294...93K} {294, 93}

\bibitem[\protect\citeauthoryear{{Landolt}}{{Landolt}}{2009}]{Land09}
{Landolt} A.~U.,  2009, \mn@doi [\aj] {10.1088/0004-6256/137/5/4186}, \href {https://ui.adsabs.harvard.edu/abs/2009AJ....137.4186L} {137, 4186}

\bibitem[\protect\citeauthoryear{{Landolt}}{{Landolt}}{2013}]{Land13}
{Landolt} A.~U.,  2013, \mn@doi [\aj] {10.1088/0004-6256/146/5/131}, \href {https://ui.adsabs.harvard.edu/abs/2013AJ....146..131L} {146, 131}

\bibitem[\protect\citeauthoryear{{Menzies}, {Cousins}, {Banfield}  \& {Laing}}{{Menzies} et~al.}{1989}]{Menzies1989}
{Menzies} J.~W.,  {Cousins} A.~W.~J.,  {Banfield} R.~M.,   {Laing} J.~D.,  1989, South African Astronomical Observatory Circular, \href {https://ui.adsabs.harvard.edu/abs/1989SAAOC..13....1M} {13, 1}

\bibitem[\protect\citeauthoryear{{Menzies}, {Marang}, {Laing}, {Coulson}  \& {Engelbrecht}}{{Menzies} et~al.}{1991}]{Menzies1991}
{Menzies} J.~W.,  {Marang} F.,  {Laing} J.~D.,  {Coulson} I.~M.,   {Engelbrecht} C.~A.,  1991, \mn@doi [\mnras] {10.1093/mnras/248.4.642}, \href {https://ui.adsabs.harvard.edu/abs/1991MNRAS.248..642M} {248, 642}

\bibitem[\protect\citeauthoryear{{Sung} \& {Bessell}}{{Sung} \& {Bessell}}{2000}]{Sung2000}
{Sung} H.,  {Bessell} M.~S.,  2000, \mn@doi [\pasa] {10.1071/AS00041}, \href {https://ui.adsabs.harvard.edu/abs/2000PASA...17..244S} {17, 244}

\makeatother
\end{thebibliography}

% Don't change these lines
\bsp	% typesetting comment
\label{lastpage}
\end{document}